# Tuning Molecule-Mediated Spin Coupling in Bottom-Up Fabricated Vanadium-TCNE Nanostructures

Daniel Wegner,[1] Ryan Yamachika,[1] Xiaowei Zhang,[1] Yayu Wang,[1] Tunna Baruah,[2] Mark R. Pederson,[3] Bart M. Bartlett,[4] Jeffrey R. Long,[4] and Michael F. Crommie[1]

[1] Department of Physics, University of California, Berkeley, and Materials Sciences Division, Lawrence Berkeley National Laboratory, Berkeley, CA 94720-7300

[2] Department of Physics, University of Texas, El Paso, TX 79968

[3] Center for Computational Materials Science, Naval Research Laboratory, Code 6390, Washington, DC 20375

[4] Department of Chemistry, University of California, Berkeley, CA 94720-1460

We have fabricated hybrid magnetic complexes from V atoms and tetracyanoethylene (TCNE) ligands via atomic manipulation with a cryogenic scanning tunneling microscope. Using tunneling spectroscopy we observe spin-polarized molecular orbitals as well as Kondo behavior. For complexes having two V atoms, the Kondo behavior can be quenched for different molecular arrangements, even as the spin-polarized orbitals remain unchanged. This is explained by variable spin-spin (i.e., V–V) ferromagnetic coupling through a single TCNE molecule, as supported by density functional calculations.

Molecule-based spintronics raises new possibilities for chemically engineering electronic and magnetic device properties with unprecedented precision at the nanoscale [1, 2]. For this concept to become reality, however, a more fundamental understanding of substrate-supported magnetic molecules is necessary since their charge, spin, and magnetic anisotropy can change as they come into contact with a surface [3, 4]. Scanning tunneling microscopy (STM) and spectroscopy (STS) offer powerful tools for studying the local magneto-electronic structure of such molecular nanostructures. Direct (atom-atom) and substrate-mediated (atom-substrate-atom) magnetic coupling between atomic spin centers has been measured previously using STM-based techniques [5-9]. Tunable spin coupling through a single molecular linker (atom-molecule-atom), however, has not yet been directly observed. One good candidate for such purposes is the strong π-electron acceptor tetracyanoethylene (TCNE). Bulk V(TCNE)$_x$ ($x \sim 2$), for example, is a molecule-based magnet with a Curie temperature $T_C \sim 400$ K [10-13].

Here we describe a bottom-up approach to create new magnetic nanostructures that display variable molecule-mediated spin-spin coupling using STM-based manipulation and characterization of individual V atoms and TCNE molecules at the surface of Ag(100). Central to this technique is our discovery that V and TCNE can be induced to form a rigid chemical bond through STM manipulation. The molecular orbital from which V-based spin is derived in the resulting complexes can be directly observed in our STS measurements. We find that for complexes consisting of one V atom and either one or two TCNE molecules, this spin is screened by the Ag substrate via the formation of a Kondo resonance. For complexes consisting of two V atoms and one TCNE molecule the Kondo effect can be switched on and off by a minute structural change that leaves the spin-containing orbital essentially unchanged. This is explained by a tunable, structure-dependent change in the V−V spin coupling strength transmitted via a single TCNE molecule. This physical picture is supported by spin-polarized density functional theory (SP-DFT) calculations that suggest that V atoms are coupled through a TCNE molecule by a tunable ferromagnetic (FM) interaction. The present findings offer a new route for designing molecular spin nanostructures with atomic-scale precision.

The experiments were performed in ultrahigh vacuum using a homebuilt STM operated at $T = 7$ K. A Ag(100) single crystal was cleaned by standard sputter-annealing procedures, and then exposed at $T = 300$ K to TCNE molecules dosed through a leak valve [14]. After *in-situ* transfer into the cryogenic STM, V atoms were deposited onto the cold sample using electron-beam evaporation. STS was performed by measuring the differential conductance $dI/dV$ as a function of the sample bias $V$ by standard lock-in techniques (modulation amplitude ~ 1 mV$_{rms}$, frequency ~ 451 Hz) under open-feedback conditions.

Typical STM images of samples prepared in this way (Fig. 1a) show isolated V atoms as round protrusions adsorbed at hollow sites on Ag(100) terraces, while coexisting TCNE molecules are identified as faint oval protrusions (surrounded by a trench) centered over top sites (Fig. 2a) [14]. While isolated V adatoms cannot be moved on Ag(100), TCNE molecules can be slid with great precision along the surface using lateral STM manipulation [15] and can thus be attached to individual V adatoms (see arrows in Fig. 1a). V-TCNE complexes formed in this way (Figs. 1b, 2b) display a merging of V and TCNE, as well as a strong change in apparent height. Once merged, the entire fused V-TCNE complex can be moved as a single unit along the surface by STM manipulation.

Larger $V_x(TCNE)_y$ structures can be built reliably by connecting TCNE or V-TCNE with other building blocks. Fig. 2c shows a complex after attachment of a second TCNE to a previously created V-TCNE complex. The newly formed $V(TCNE)_2$ complex again shows a rigid connection of both TCNE molecules to the interior V atom. Linear $V_2TCNE$ complexes can be made by connecting a V-TCNE molecule to a V adatom (Figs. 1b, 2d,e). STM images of these different bottom-up synthesized molecules are shown in Fig. 2, along with structural models based on analysis of the images. In all cases the fused V atoms stay close to four-fold hollow sites on the Ag(100) surface, but the fused TCNE prefers bridge sites, thereby reducing the V-N distance. Linear $V_2TCNE$ structures can be built reproducibly in two different forms that are distinguishable by the V−V angle relative to the [001]-direction of the substrate: the first structure ($V_2TCNE@27°$) exhibits an angle of 27 ± 1° (Fig. 2d), while the second structure ($V_2TCNE@11°$) has an angle of 11 ± 1° (Fig. 2e). These two isomers exhibit an overall difference in size, the distance between V atoms being slightly larger (by ~ 1 Å) in the $V_2TCNE@27°$ molecule as compared to the shorter $V_2TCNE@11°$.

In order to understand the magneto-electronic behavior of the newly synthesized $V_x(TCNE)_y$ complexes, we performed STS experiments on these structures with sub-nanometer resolution (Figs. 3 and 4). We observed three different features in this spectroscopy: (*i*) molecular orbital resonances, (*ii*) vibrational inelastic features, and (*iii*) Kondo resonances. Molecular orbital resonances are seen at "high" biases ($|V| > 100$ mV) while inelastic features are seen at "low" biases ($|V| < 50$ mV), and Kondo resonances are observed to straddle the Fermi energy at $V = 0$. The inelastic features reveal the nature of the V-TCNE bonding while the molecular orbital resonances reveal the presence of spin in the $V_x$TCNE$_y$ complexes. The Kondo resonance observations reveal how the molecular spins are coupled to their environment, including adjacent spins.

We first describe the inelastic features (marked with the label "$E_{vib}$" in Fig. 3). These are seen as steps in $dI/dV$ or peaks (dips) at positive (negative) voltage in $d^2I/dV^2$, and they arise as new tunneling channels are opened due to molecular excitations [16]. For bare TCNE (Fig. 3d) a very clear inelastic mode exists at ~ 30 mV. We identify this as the TCNE rocking or wagging mode that is known to lie at this energy from optical spectroscopy measurements and DFT calculations [17, 18]. When a single V atom is attached to a single TCNE molecule, however, a new mode appears at 45 mV that is localized to the V-site (Fig. 3e). We identify this as the V-N stretch vibration, and its energy corresponds well to the V-N stretch mode measured in other structures by optical spectroscopy [19]. The presence of the V-N stretch vibration provides strong evidence that V atoms connected to TCNE via STM manipulation are covalently bonded. When two TCNE molecules are connected to one V atom, the V-N stretch mode is no longer detectable, but the TCNE rocking/wagging mode is still seen over the TCNE molecules (Fig. 3f).

We now describe the orbital resonances seen at larger biases for $V_x$TCNE$_y$ complexes. In the data of Figs. 3,4 these resonances are marked by the label "$E_d$", since they are believed to arise from V $d$-orbitals. STS of V-TCNE complexes (Fig. 3b) shows a pronounced broad molecular resonance at $E_d = -0.17$ V that has strong amplitude on the V-site and slightly lower amplitude on the TCNE region of the complex. This resonance does not exist for bare TCNE on Ag(100) [14]. When two TCNE molecules are attached to a single V atom, the orbital resonance shifts to a slightly reduced energy $E_d$ = -0.25 V (Fig. 3c) and becomes localized to the V atom and cannot be seen over adjacent TCNE molecules. When two V atoms are bonded to a single TCNE molecule (as shown in Fig. 2d,e), the molecular orbital shifts upward slightly to $E_d$ = -0.15 V and has amplitude on both the V and TCNE regions of the complex. This behavior is seen identically for both $V_2$TCNE@27° and $V_2$TCNE@11° (Fig. 4a,b).

We now describe our observations of Kondo resonances in the low bias spectra of $V_x$TCNE$_y$ complexes. A Kondo resonance is the spectral signature of a many-body electron cloud that occurs as the itinerant spins of a nonmagnetic metal screen the spin of a local magnetic moment [20, 21]. V-TCNE exhibits a Kondo resonance having amplitude both on the V atom and the adjacent TCNE molecule (Fig. 3e) with a width of $\Gamma$ = 11 ± 2 mV, indicating a Kondo temperature of $T_K$ ~ 65 K. Figure 3f shows that the Kondo resonance remains for the V(TCNE)$_2$ complex ($\Gamma$ = 6 ± 1 mV, i.e., $T_K$ ~ 35 K), except that it now has amplitude only on the V atom. The spatial dependence of the Kondo resonance for both V-TCNE and V(TCNE)$_2$ thus mirrors the spatial dependence of the molecular orbital at $E_d$. When two V atoms are connected to a single TCNE molecule in the $V_2$TCNE@27° configuration the Kondo resonance also remains (in this case, $\Gamma$ = 10 ± 2 mV, i.e., $T_K$ ~ 60 K), and again the spatial distribution of the Kondo resonance mirrors that of the $E_d$ orbital resonance. Surprisingly, however, when two V atoms are connected to a TCNE molecule in the $V_2$TCNE@11° configuration, there is no Kondo resonance to be seen on either the V or TCNE sites (Fig. 4d) despite the fact that the $E_d$ orbital resonance amplitude remains unchanged across the entire V$_2$TCNE@11° complex.

This behavior can be explained by identifying the filled orbital resonance at $E_d$ as the signature of a local magnetic moment. The presence (or absence) of the Kondo resonance then indicates how this local moment couples to its environment. In the case of V$_2$TCNE our data indicates that, although nearly identical local spin moments exist for V$_2$TCNE@27° and V$_2$TCNE@11°, the V−V spin coupling is very different for these two complexes. The V spins in V$_2$TCNE@11° appear to be FM coupled through the TCNE molecule with a larger exchange energy than for the more weakly coupled V$_2$TCNE@27° complex. The signature of this tunable V−V exchange coupling is the disappearance of the Kondo effect in the more strongly coupled V$_2$TCNE@11° case [22]. This interpretation of the data is supported by SP-DFT calculations as well as by differences in V-geometry between V$_2$TCNE@27° and V$_2$TCNE@11°, as described below.

First-principles SP-DFT calculations of the magneto-electronic structure of isolated V$_x$TCNE$_y$ complexes were performed using the NRLMOL code [23-26]. All atoms in the calculation were treated within an LCAO formulation at the all-electron level. The basis sets used here are roughly equivalent to triple-$\zeta$ or better. The generalized gradient approximation (PBE-GGA) was used to approximate the exchange-correlation functional [27]. Structures were relaxed until all forces were below 0.001 Hartree/Bohr. Figure 3a shows the local density of states (LDOS) calculated for a V-TCNE complex. The majority-spin LDOS shows a pronounced filled state at $E_d$ = -0.20 eV, implying that the experimental orbital resonance seen at this same energy is a majority-spin state arising from a V $d$-orbital. Calculations of a V$_2$TCNE complex (Fig.

4b, inset) also show a dominating V $d$-resonance at $E_d$ = -0.12 eV, similar to what is seen for the experimental case (Fig. 4a,b). This agreement between theory and experiment only occurs when the simulated V-TCNE and $V_2$TCNE complexes are charged with one electron, which we assume is drawn from the Ag substrate in the experiment [28]. In this case the FM state is lower in energy than the antiferromagnetic state by 170 meV. The SP-DFT calculations thus support the conclusion that orbital resonances seen experimentally at $E_d$ indicate the presence of a local moment, and that these moments are FM coupled for anionic $V_2$TCNE.

The reason that the Kondo effect is quenched for $V_2$TCNE@11° and not for $V_2$TCNE@27° is because the 11° complex is more strongly FM coupled through the TCNE molecule than the 27° complex. Although our SP-DFT calculations are not accurate enough to distinguish between these two cases, we can see from the structure of the 11° complex (Fig. 2e) that the V atoms are closer together than for the 27° complex (Fig. 2d). This couples them more strongly to the TCNE ligand, thus providing stronger FM coupling. As this coupling strength rises above the single-moment Kondo temperature, the two V atoms form a FM complex rather than two individually screened magnetic moments. FM coupled spin complexes are well known to have a lower Kondo temperature than single impurities, thus explaining why the Kondo effect is quenched for the 11° complex [22]. The fact that the binding energies of the spin-containing orbital states at $E_d$, as well as their widths (i.e. their hybridization with the substrate), are identical for both $V_2$TCNE@11° and $V_2$TCNE@27° provides further evidence that quenching of the Kondo effect comes from tunable ligand-induced FM exchange coupling rather than a simple shifting of the single-impurity Kondo temperature.

We emphasize that the separation between the two V atoms in V$_2$TCNE@11° (~10 Å) is relatively large. Strong FM coupling over such distances is possible only due to molecular ligand-induced exchange coupling [5, 8]. This direct use of a molecule as a mediating unit between spin centers enables a new type of spin-coupling engineering. For example, the influence of molecular size on magnetic coupling might be studied systematically by replacing TCNE with closely related molecules (e.g. TCNQ), or the effects of chemical interactions might be studied by varying functional end groups. Such a bottom-up strategy opens new paths for designing quantum spin structures with atomic-scale precision.

This work was supported by NSF NIRT Grant ECS-0609469. D.W. is grateful for funding by the Alexander von Humboldt Foundation. Y.W. thanks the Miller Institute for a research fellowship. T.B. gratefully acknowledges the computer resources at Naval Research Laboratory and at UTEP.

**Figure Captions:**

**Fig. 1:** (a) STM image of TCNE molecules and V atoms adsorbed on Ag(100). (b) Same area after moving TCNE molecules toward V atoms using STM manipulation [following arrows in (a)], thus forming V-TCNE complexes. The arrow indicates how a V-TCNE complex will be moved toward a second V atom to form a $V_2$TCNE complex (see Fig. 2d,e). Both STM images are plotted with the same scale bar for apparent height. Imaging parameters: 1 V, 5 pA.

**Fig. 2:** Highly resolved STM images and structural models of (a) TCNE, (b) V-TCNE, (c) V(TCNE)$_2$, (d) $V_2$TCNE@27°, and (e) $V_2$TCNE@11° on Ag(100). The models are derived from the STM images. Note that $V_2$TCNE@11° has a shorter V-V distance compared to $V_2$TCNE@27°. All STM images are plotted with the same height scale bar shown in Fig. 1. Imaging parameters: 1 V, 5-50 pA.

**Fig. 3:** (a) SP-DFT-calculated spin-resolved LDOS of a free anionic V-TCNE complex. Large-bias STS spectrum is shown for (b) V-TCNE and (c) V(TCNE)$_2$. Highly resolved low-bias STS is shown for (d) TCNE, (e) V-TCNE, and (f) V(TCNE)$_2$. For clarity, the Ag-background signal (green curve in d) was subtracted from spectra in (b), (c), (e), and (f).

**Fig. 4:** Large-bias (a,b) and highly resolved low-bias (c,d) STS spectra of $V_2$TCNE@27° and $V_2$TCNE@11°, respectively (Ag background spectrum is subtracted). The inset in (b) shows the calculated spin-resolved LDOS of a free anionic $V_2$TCNE complex.

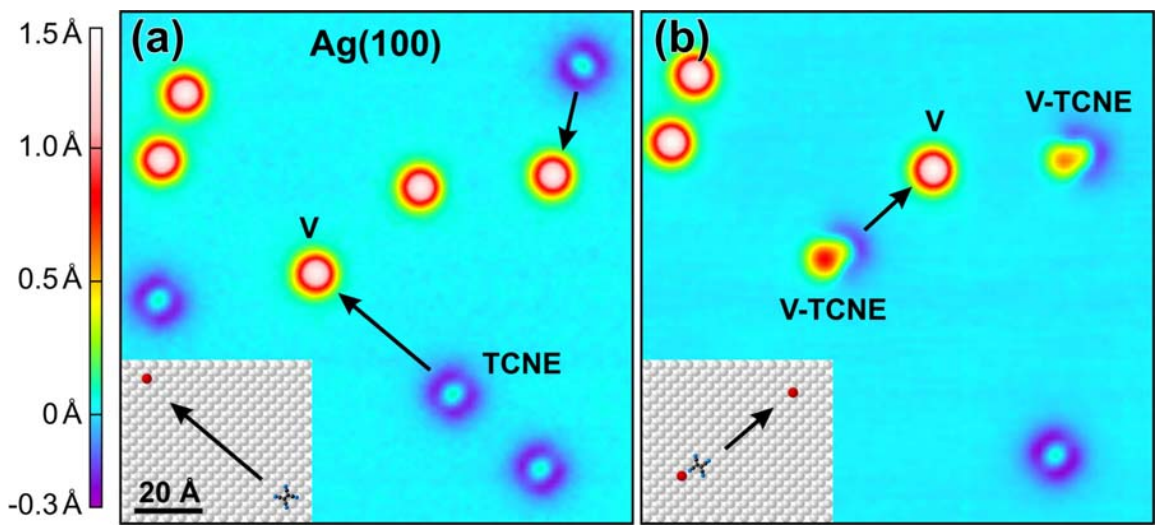

Figure 1

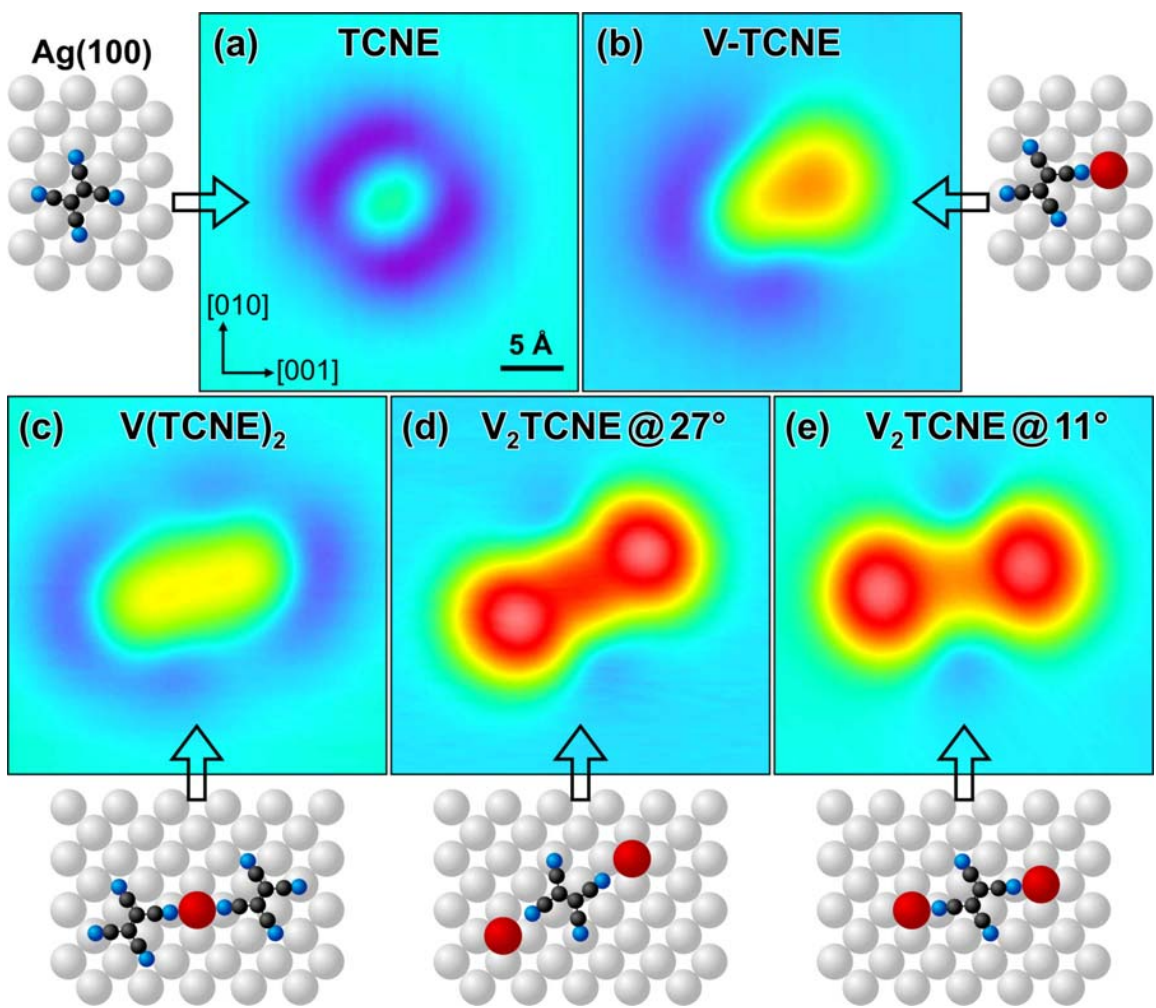

Figure 2

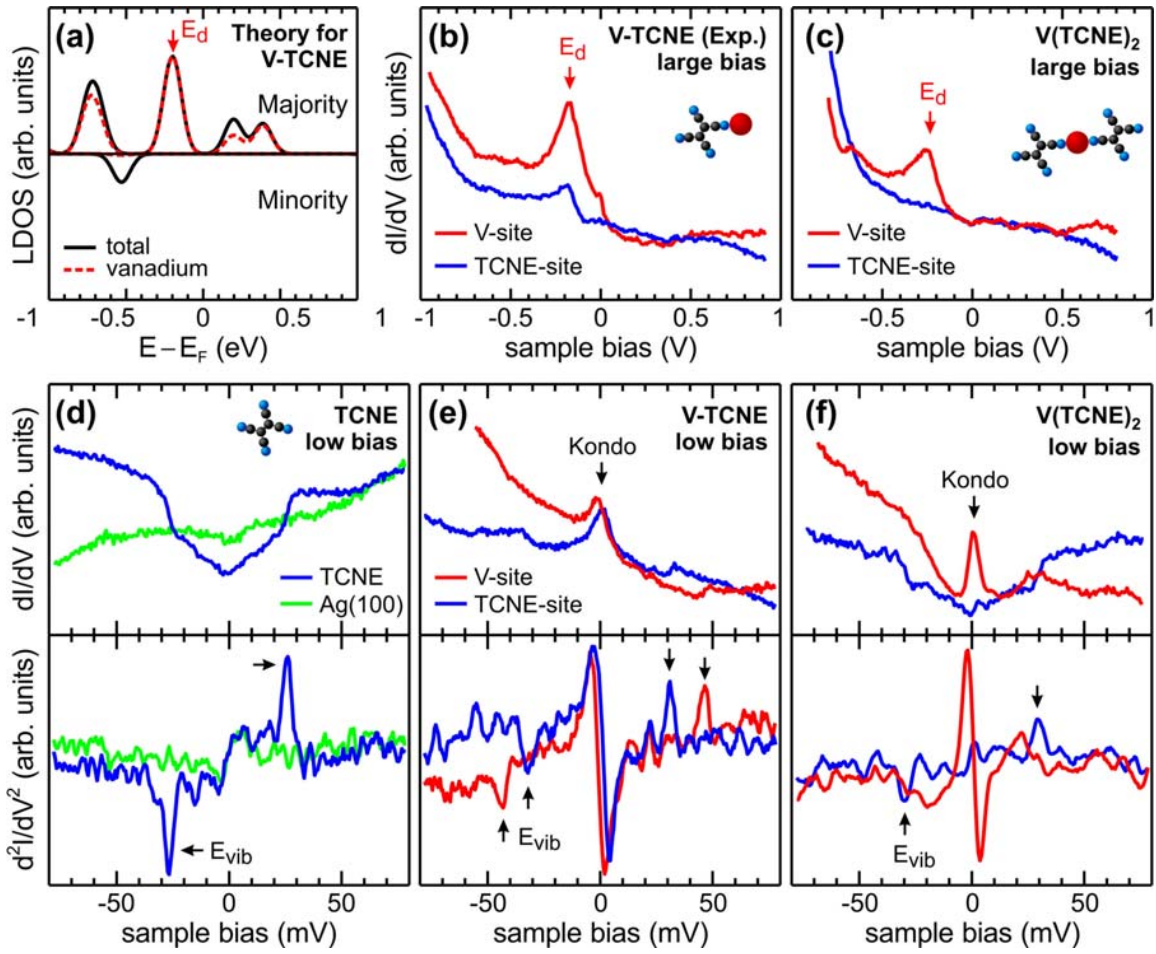

Figure 3

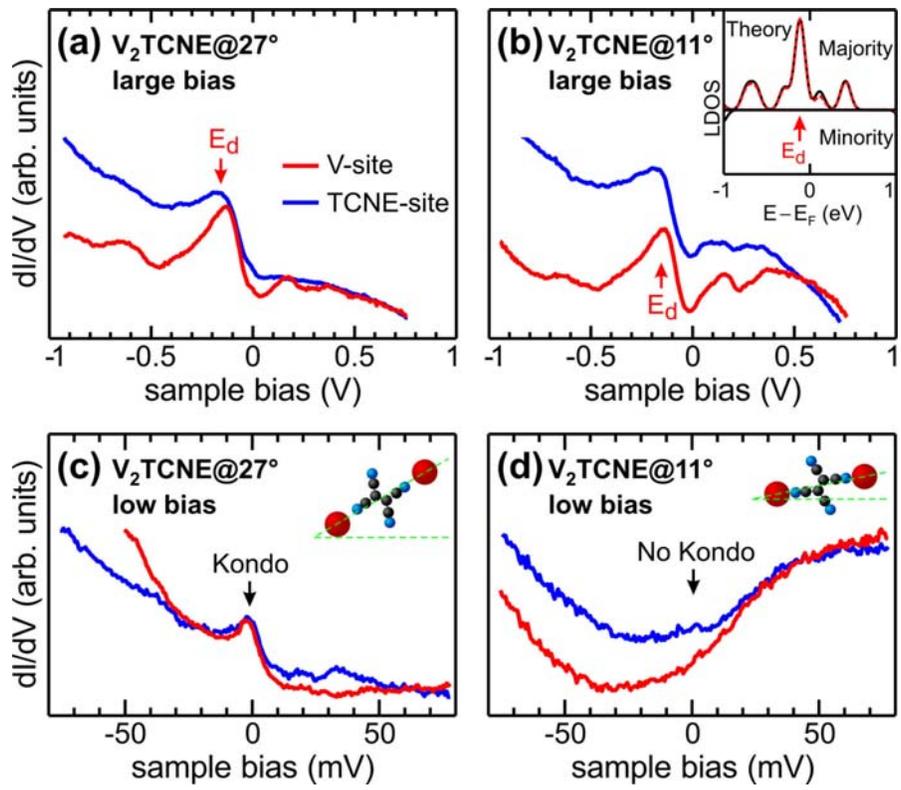

Figure 4